\documentclass[pre, aps, reprint, superscriptaddress]{revtex4-2}

\usepackage[table,xcdraw]{xcolor}
\usepackage{amsfonts}
\usepackage{amsmath}
\usepackage{mathtools}
\usepackage{float}
\usepackage{color}
\usepackage{gensymb}
\usepackage{multirow}
\usepackage{makecell}
\usepackage[mathscr]{euscript}
\usepackage{pifont}
\usepackage{mathtools}

\newcommand{\e}[1]{\ensuremath{\times 10^{#1}}}
\newcommand{\ee}[1]{\ensuremath{10^{#1}}}
\newcommand{\relu}[0]{\ensuremath{\mbox{RELU}}}

\usepackage{empheq}
\usepackage{physics}
\newcommand{\ignore}[1]{}

\begin{document}

\title{Correlation between entropy and generalizability in a neural network}

\author{Ge Zhang}
\affiliation{\emph{Department of Physics}, \emph{City University of Hong Kong}, Hong Kong, China}

\begin{abstract}

Although neural networks can solve very complex machine-learning problems, the theoretical reason for their generalizability is still not fully understood. Here we use Wang-Landau Mote Carlo algorithm to calculate the entropy (logarithm of the volume of a part of the parameter space) at a given test accuracy, and a given training loss function value or training accuracy. Our results show that entropical forces help generalizability. Although our study is on a very simple application of neural networks (a spiral dataset and a small, fully-connected neural network), our approach should be useful in explaining the generalizability of more complicated neural networks in future works.

\end{abstract}

\maketitle

\section{Introduction}

Neural networks has gained tremendous success in solving machine-learning problems since the past decade or so. One remaining challenge in this field, though, is theoretically understanding their success. Conceptually, many neural-network applications are similar to fitting. We know that in traditional polynomial fitting, if the polynomial has too many degrees of freedom, then it tends to overfit the noise in the data, resulting in poor predictions. Neural networks contain way more degrees of freedom, so why don't they overfit?

The above question is one of the central issue in machine learning, and this paper certainly does not offer a complete answer. Here we demonstrate that a statistical-physical approach is valuable when searching for answers. We study the correlation between the entropy and generalizability when a neural network well fits a training data set (low training loss or high training accuracy). Here the entropy is defined as the logarithm of the volume in the high-dimensional parameter space that correspond to a particular training error, training loss, or test error. We calculate the entropy using Wang-Landau Monte Carlo, and show that for a simple machine-learning problem (a fully-connected neural network applied to the spiral dataset), at zero training error or at a very low training loss, the entropically-favored states have very high test accuracies that increases as the network depth increases. Our results suggest that while parameter choices with serious overfitting problems exist, they simply have too low entropy to be a probable result after training.

\section{Definitions}
In traditional polynomial fitting, one defines a parameterized function, e.g., 
\begin{equation}
    f(\mathbf c, x)=c_1+c_2x+\cdots+c_N x^{N-1},
\end{equation}
where $\mathbf c=(c_1, c_2, \cdots, c_N)$ is a vector representation of the parameters and $x$ is the independent variable; and finds parameters that fits a dataset. The dataset consists of pairs of values $(x_i, y_i)$, where $i=1, 2, \cdots, N_d$, $N_d$ is the number of data points, and $y_i$ are the expected function value. The parameters are usually found by minimizing a function that quantifies the difference between the prediction and data, for example
\begin{equation}
    L(\mathbf c)=\sum_{i=1}^{N_d} \left[f(\mathbf c, x_i)-y_i \right]^2.
    \label{eq:loss1}
\end{equation}
$L(\mathbf c)$ is called the ``loss function'' in the machine-learning community.

A neural network is conceptually similar but has the following differences. First, the function input $\mathbf x$ is usually a vector rather than a scalar. Second, the parameterized funcion is usually way more complicated. Here we consider a relatively simple case, a fully-connected neural network with $H$ hidden layers of $W$ neurons. Let $\mathbf a_i$ be the intermediate result after the $i$th layer, and let $\mathbf a_0=\mathbf x$. 
The dimensionality of $\mathbf a_1, \mathbf a_2, \cdots, \mathbf a_{H}$ are a constant $W$. Both $\mathbf a_{0}$ and $\mathbf a_{H+1}$ are two-dimensional vectors in this paper, in order to suit the problem we will study (detailed later). 
Each layer performs the following computation
\begin{equation}
    \mathbf a_i=\relu(w_i \mathbf a_{i-1}+\mathbf b_i),
\end{equation}
where $w_i$ are matrices, the number of rows and columns of $w_i$ are chosen so that the multiplication with $\mathbf a_{i-1}$ produces the desired dimensionality for $\mathbf a_{i}$, $b_i$ are vectors, and $\relu()$ function changes each negative component of a vector to zero while leaving the positive components unchanged. All elements of $w_i$ and $b_i$ are considered fitting parameters. In other words, $\mathbf c$ is a very long vector defined as the concatenation of $w_i$ and $b_i$ for all $i$.

Third, while the prediction from a polynomial fit is a continuous variable, many neural-network applications desires categorical predictions. In such case, the most common practice is to append a so-called ``softmax layer'' to the neural network
\begin{equation}
    a_{H+2, i}=\frac{\exp(a_{H+1, i})}{\sum_i \exp(a_{H+1, i})}.
    \label{eq:prediction}
\end{equation}
where $a_{H, i}$ is the $i$th component of vector $\mathbf a_H$. Since Eq.~(\ref{eq:prediction}) guarantees that $\sum_i a_{H+2, i}=1$, one can then interpret $a_{H+2, i}$ as the probability that the input belongs to the $i$th category. 
The accuracy $A$ is defined as the probability that $a_{H+2, y_i}$ is the largest element of vector $\mathbf a_{H+2}$.
The loss function, Eq.~(\ref{eq:loss1}), is also changed to become suitable for categorical predictions
\begin{equation}
    L(\mathbf c)=-\sum_i^{N_d} \ln{a_{H+2, y_i}}.
\end{equation}
The idea is that for the $i$th data point, since $y_i$ is the desired prediction, the higher $a_{H+2, y_i}$ is, the more confident the neural network is on the correct answer, and the lower $L$ is. Machine-learning researchers found that adding a so-called ``regularization term'' to $L$ can often improve generalization, so we will also try that. The resulting loss function is
\begin{equation}
    L(\mathbf c)=-\sum_i^{N_d} \ln{a_{H+2, y_i}}+\lambda |\mathbf c|^2,
\end{equation}
where $\lambda$ is a constant called the ``regularization parameter.''

\begin{figure}
\includegraphics[width=0.49\textwidth]{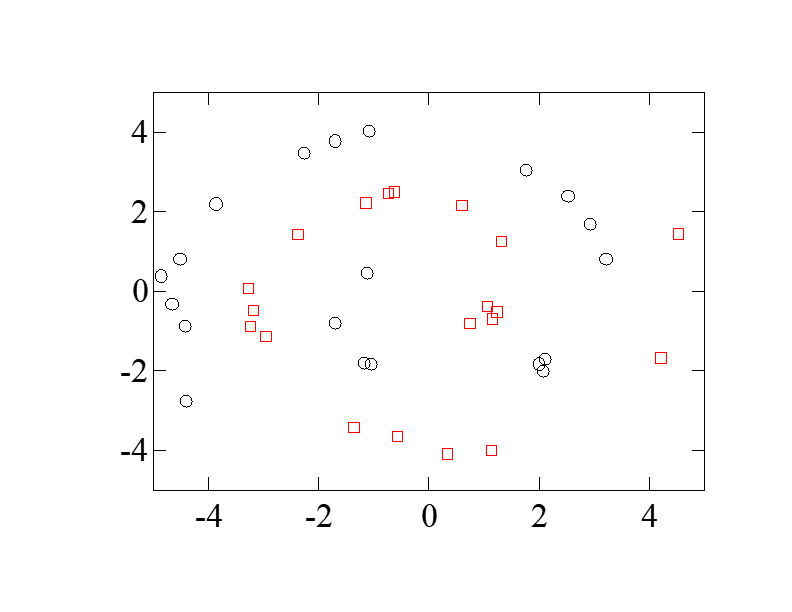}
\caption{One example of our dataset.
}
\label{fig:spiral}
\end{figure}

The neural network is applied on the so-called ``spiral'' dataset, one example is shown in Fig.~\ref{fig:spiral}. Black dots and red dots form two arms of a spiral in a 2D plane. Given the horizontal and vertical coordinates of a dot, the neural network is asked to predict the color of the dot. More specifically, we compute the $i$th datapoint with the following formula:
\begin{equation}
    x_i=(r_i \cos\theta_i+N_{1i}, r_i \sin\theta_i+N_{2i}),
\end{equation}
where $r_i$ is uniformly distributed between 1 and 5, $\theta_i=2r_i+\pi y_i$, and both $N_{1i}$ and $N_{2i}$ are Gaussian random noises with mean 0 and standard deviation 0.1. We generate 20 random black dots ($y_i=0$), and 20 random red dots ($y_i=1$). They constitute our training set. We then generate a test set following exactly the same distribution. 

We define the set of all possible choices of $\mathbf c$ as the ``parameter space.'' The parameter space has an infinite volume, but we will study a finite-volume region. Machine-learning practitioners found that training the neural-network is most efficient when each element of the weight matrix is on the order of $1/\sqrt{W}$. We therefore limit each component of $\mathbf c$ to the range $[-2/\sqrt{W}, 2/\sqrt{W}]$. We will then discuss specific parts within this region, e.g., the part that correspond to $100\%$ training accuracy and $90\%$ test accuracy. The entropy for each part of the parameter space is defined as
\begin{equation}
    S=\ln(V_p/V_{\mbox{all}}),
\end{equation}
where $V_p$ is the volume of this part of the parameter space, $V_{\mbox{all}}=\left( 4/\sqrt{W}\right)^d$ is the total volume of the region of parameter space we are studying, and $d$ is the number of parameters.

\section{Numerical procedure}
We use Wang-Landau Monte Carlo \cite{wang2001efficient} to compute such entropies. Our results are averaged over 6 realizations of the training and test datasets in Sec.~\ref{sec:result1} and 24 realizations of the datasets in Sec.~\ref{sec:result2}. For each pair of datasets, we perform $\ee{4}$ stages of Monte Carlo simulation. Each stage contains $3.2\e{6}$ trial moves. The step sizes of the trial moves are adjusted on-the-fly so that probability of accepting is between 0.3 and 0.7: after every 1000 trail moves, the step size is increased or decreased by 10\% if the acceptance rate is too high or too low. The ``modification factor'' in Ref. \cite{wang2001efficient} is $f=\exp[5/(n+10)]$, for the $n$th stage.

\section{Results}
\subsection{calculating entropy as a function of train loss and test accuracy}
\label{sec:result1}

\begin{figure}
\includegraphics[width=0.3\textwidth]{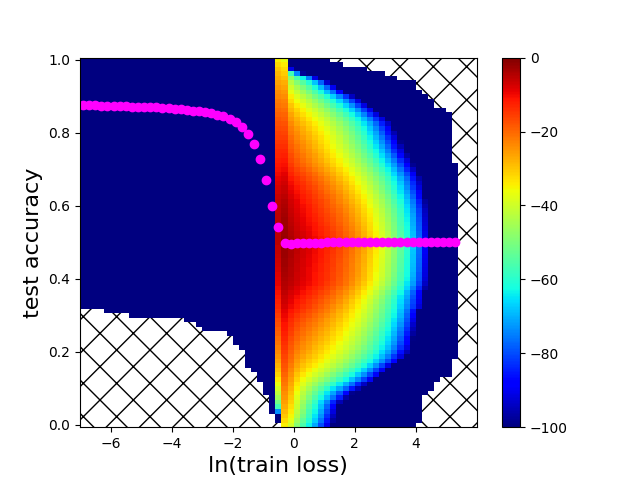}
\includegraphics[width=0.3\textwidth]{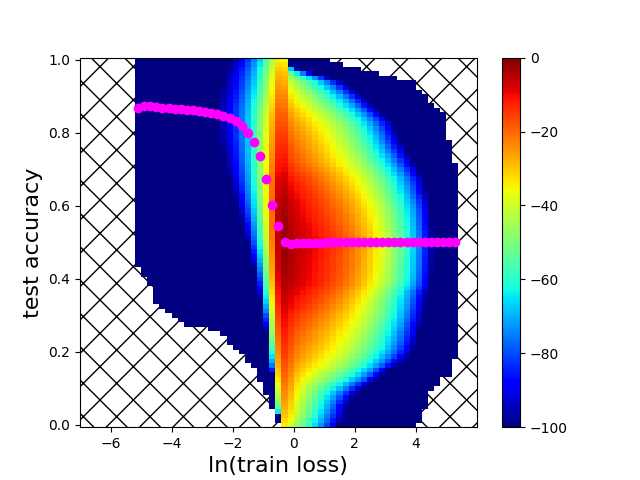}
\includegraphics[width=0.3\textwidth]{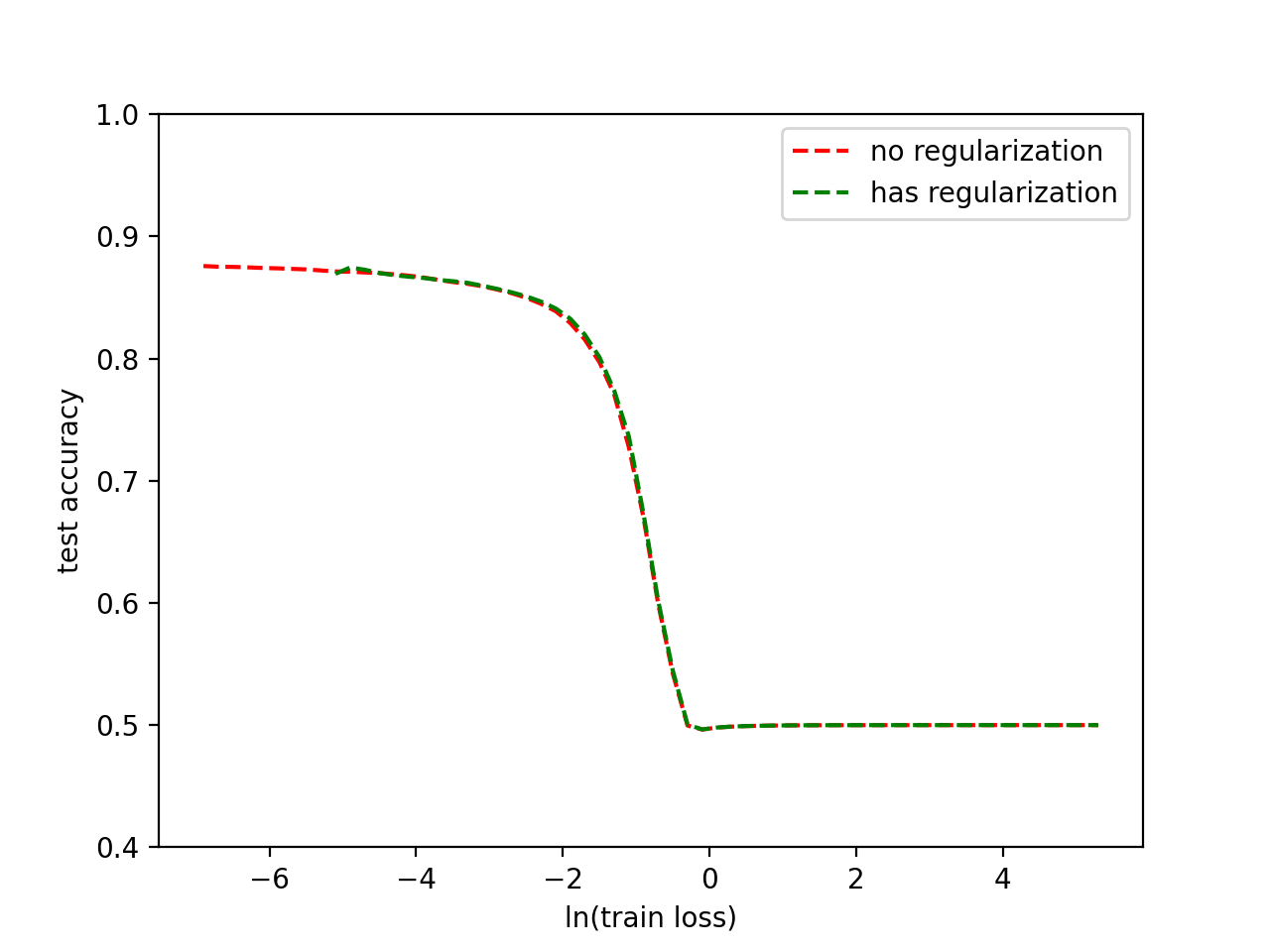}
\caption{(top) Entropy as a function of the test accuracy and the loss function on the training dataset without regularization. At each train loss value, we compute the average test accuracy of an equilibrium state (see main text for details) and mark it using a magenta dot. (middle) Same as top, except with regularization parameter $\lambda=\ee{-4}$. (bottom) Comparing the equilibrium test accuracy for the two cases.
}
\label{fig:sLoss}
\end{figure}

We can calculate the entropy as a function of the training loss and the test accuracy, $S(L_{\mbox{train}}, A_{\mbox{test}})$. Results for $H=4$ and $W=8$ are shown in Fig.~\ref{fig:sLoss}.
For each bin of $L_{\mbox{train}}$, we can calculate the equilibrium test accuracy, i.e., the test accuracy averaged over their weight determined by the entropy
\begin{equation}
    \langle A_{\mbox{test}}(L_{\mbox{train}}) \rangle=\frac{\int_0^1 x \exp\left[S(L_{\mbox{train}}, x)\right] dx}{\int_0^1 \exp\left[S(L_{\mbox{train}}, x)\right] dx}.
\end{equation}
Such test accuracies are plotted as magenta dots in Fig.~\ref{fig:sLoss}. We also compare $\langle A_{\mbox{test}}(L_{\mbox{train}}) \rangle$ curves for the cases with and without regularization. We see that at low train loss values, regularization indeed raises the test accuracy.

Having found the equilibrium test accuracy, we can compare it with the test accuracy of typically trained neural networks. The comparison is presented in Table~\ref{tab:compare}. For each set of conditions (each row) we trained 100 instances of the neural network and presented their average test accuracy here. We train our neural networks using the ``Adam optimizer'' \cite{kingma2014adam} with learning rate $\ee{-3}$. We did not divide the training data into batches. For all cases, the trained neural network possessed test accuracies that are much worse than the equilibrium state. This shows that a typically trained neural network is in an non-equilibrium state, with generalization performance worse than the equilibrium state at the same $L_{\mbox{train}}$.

\begin{table}[]
\caption{Comparison of the test accuracy of a neural network trained with gradient descent algorithm and the test accuracy of the equilibrium state.}
\begin{tabular}{c|c|c|c|c|c}
 $H$ & $W$ &  \thead{training time \\ (epochs)} & $L_{\mbox{train}}$ & \thead{ trained \\ $A_{\mbox{test}}$} & \thead{equilibrium \\ $A_{\mbox{test}} $}\\ \hline
3                & 6                 & 2000                   & 0.152         & 71\%                 & 83\%                      \\
3                & 8                 & 2000                   & 0.077         & 73\%                 & 85\%                      \\
4                & 6                 & 2000                   & 0.103         & 73\%                 & 84\%                      \\
4                & 8                 & 2000                   & 0.045         & 74\%                 & 86\%                      \\
3                & 6                 & 5000                   & 0.082         & 74\%                 & 85\%                      \\
3                & 8                 & 5000                   & 0.016         & 77\%                 & 87\%                      \\
4                & 6                 & 5000                   & 0.012         & 75\%                 & 85\%                      \\
4                & 8                 & 5000                   & 0.006         & 78\%                 & 87\%                     
\end{tabular}
\label{tab:compare}
\end{table}

\subsection{calculating entropy as a function of train accuracy and test accuracy}
\label{sec:result2}

\begin{figure}
\includegraphics[width=0.49\textwidth]{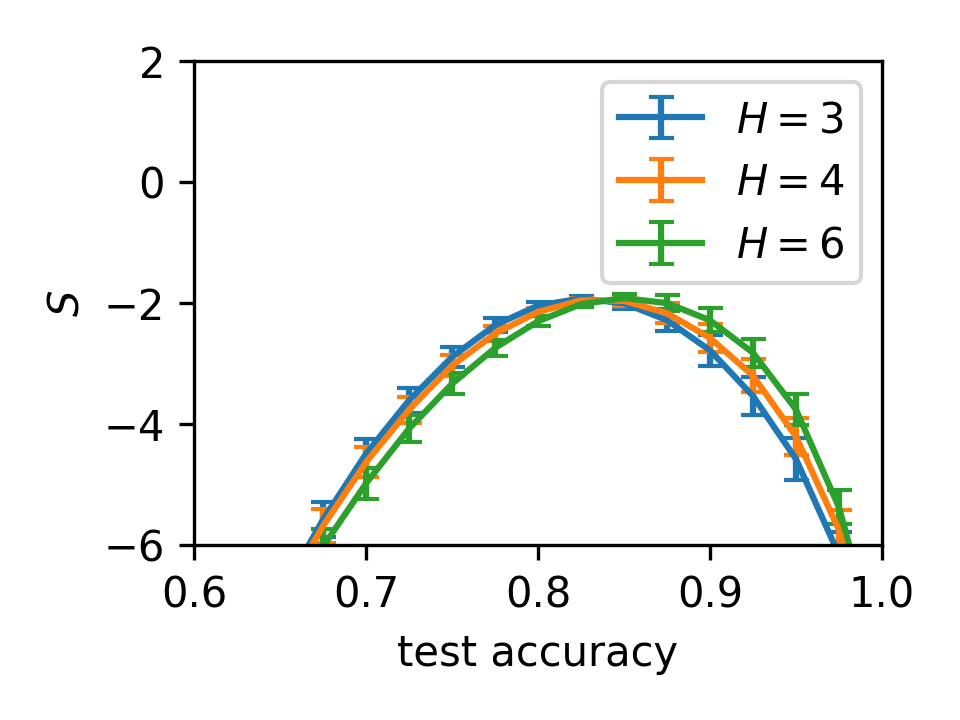}
\includegraphics[width=0.49\textwidth]{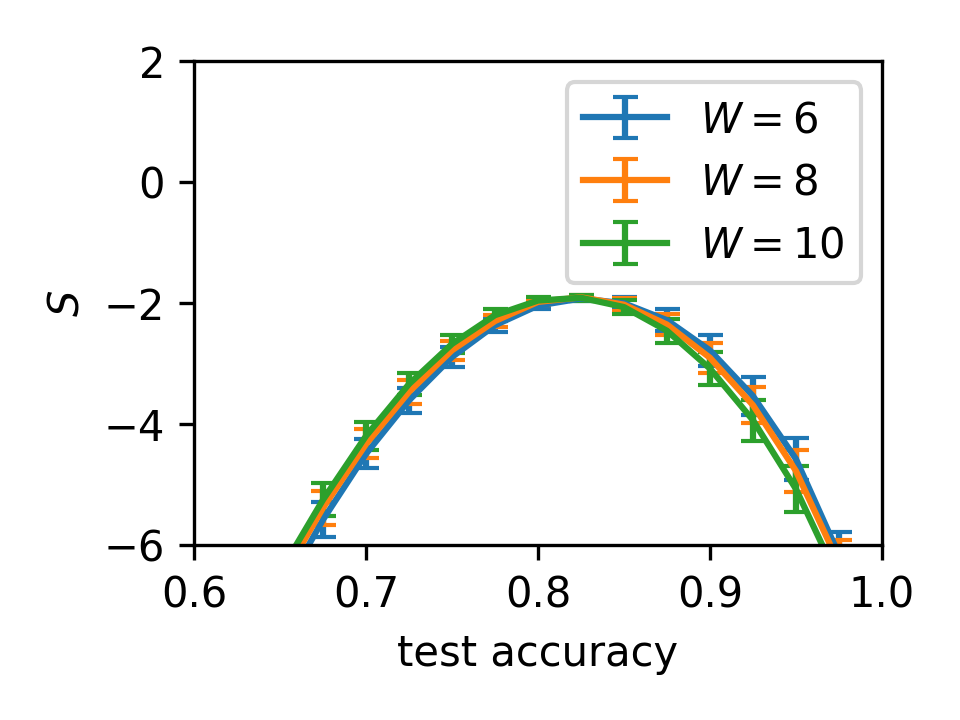}
\caption{Entropy as a function of test accuracy under the constraint that the training accuracy is 100\%. (top)  Results for $W=6$ and various $H$. (bottom) Results for $H=3$ and various $W$.
}
\label{fig:sAccuracy}
\end{figure}

We can also calculate the entropy as a function of the accuracy on the training and test datasets, $S(A_{\mbox{train}}, A_{\mbox{test}})$. We can study the relation between $S$ and $A_{\mbox{test}}$ when $A_{\mbox{train}}$ is set to be 100\%, i.e., when the neural network perfectly fits the training data. This approach appears to be more straightforward, but we lose the ability to incorporate regularization into the picture because $S$ no longer depends on any loss function. Our results are presented in Fig.~\ref{fig:sAccuracy}. We see that when the number of neurons per layer $W$ is fixed, increasing the number of layers $H$ makes the curve shift to the right. This indicates that as the neural network becomes deeper, its entropically-favored state gains generalizability. When we fix $H$ and increase $W$, however, the opposite happened. Indicating that wider neural networks possess worse entropically-favored state. The last result is counter-intuitive, and the cause should be investigated in the future. 

\section{Discussion}
In this paper, we investigated the relationship between entropy and the training loss function, training accuracy, and test accuracy. We find that an equilibrium state (roughly speaking, the max-entropy state) of a neural network possesses a much higher test accuracy than a typically trained state. We also show that the equilibrium test accuracy increases as the network depth increases.

We plan to carry out this study for more complicated neural networks and datasets, e.g., convolutional neural networks on image-classification tasks, or transformer neural networks on language/reasoning tasks. These networks are much more computationally intensive, and we plan to change our method to Wang-Landau molecular dynamics to improve efficiency. This is because molecular-dynamics-based algorithms gains better efficiency by utilizing the derivative of the loss function, which is easily available from machine-learning software.

\bibliography{references}

\begin{thebibliography}{2}%
\makeatletter
\providecommand \@ifxundefined [1]{%
 \@ifx{#1\undefined}
}%
\providecommand \@ifnum [1]{%
 \ifnum #1\expandafter \@firstoftwo
 \else \expandafter \@secondoftwo
 \fi
}%
\providecommand \@ifx [1]{%
 \ifx #1\expandafter \@firstoftwo
 \else \expandafter \@secondoftwo
 \fi
}%
\providecommand \natexlab [1]{#1}%
\providecommand \enquote  [1]{``#1''}%
\providecommand \bibnamefont  [1]{#1}%
\providecommand \bibfnamefont [1]{#1}%
\providecommand \citenamefont [1]{#1}%
\providecommand \href@noop [0]{\@secondoftwo}%
\providecommand \href [0]{\begingroup \@sanitize@url \@href}%
\providecommand \@href[1]{\@@startlink{#1}\@@href}%
\providecommand \@@href[1]{\endgroup#1\@@endlink}%
\providecommand \@sanitize@url [0]{\catcode `\\12\catcode `\$12\catcode
  `\&12\catcode `\#12\catcode `\^12\catcode `\_12\catcode `\%12\relax}%
\providecommand \@@startlink[1]{}%
\providecommand \@@endlink[0]{}%
\providecommand \url  [0]{\begingroup\@sanitize@url \@url }%
\providecommand \@url [1]{\endgroup\@href {#1}{\urlprefix }}%
\providecommand \urlprefix  [0]{URL }%
\providecommand \Eprint [0]{\href }%
\providecommand \doibase [0]{https://doi.org/}%
\providecommand \selectlanguage [0]{\@gobble}%
\providecommand \bibinfo  [0]{\@secondoftwo}%
\providecommand \bibfield  [0]{\@secondoftwo}%
\providecommand \translation [1]{[#1]}%
\providecommand \BibitemOpen [0]{}%
\providecommand \bibitemStop [0]{}%
\providecommand \bibitemNoStop [0]{.\EOS\space}%
\providecommand \EOS [0]{\spacefactor3000\relax}%
\providecommand \BibitemShut  [1]{\csname bibitem#1\endcsname}%
\let\auto@bib@innerbib\@empty
\bibitem [{\citenamefont {Wang}\ and\ \citenamefont
  {Landau}(2001)}]{wang2001efficient}%
  \BibitemOpen
  \bibfield  {author} {\bibinfo {author} {\bibfnamefont {F.}~\bibnamefont
  {Wang}}\ and\ \bibinfo {author} {\bibfnamefont {D.~P.}\ \bibnamefont
  {Landau}},\ }\bibfield  {title} {\bibinfo {title} {Efficient, multiple-range
  random walk algorithm to calculate the density of states},\ }\href
  {https://doi.org/10.1103/PhysRevLett.86.2050} {\bibfield  {journal} {\bibinfo
   {journal} {Phys. Rev. Lett.}\ }\textbf {\bibinfo {volume} {86}},\ \bibinfo
  {pages} {2050} (\bibinfo {year} {2001})}\BibitemShut {NoStop}%
\bibitem [{\citenamefont {Kingma}\ and\ \citenamefont
  {Ba}(2014)}]{kingma2014adam}%
  \BibitemOpen
  \bibfield  {author} {\bibinfo {author} {\bibfnamefont {D.~P.}\ \bibnamefont
  {Kingma}}\ and\ \bibinfo {author} {\bibfnamefont {J.}~\bibnamefont {Ba}},\
  }\bibfield  {title} {\bibinfo {title} {Adam: A method for stochastic
  optimization},\ }\href@noop {} {\bibfield  {journal} {\bibinfo  {journal}
  {arXiv preprint arXiv:1412.6980}\ } (\bibinfo {year} {2014})}\BibitemShut
  {NoStop}%
\end{thebibliography}%

\end{document}